\documentstyle[preprint,aps]{revtex}
\tighten
\draft
\widetext
\input epsf
\preprint{HUTP-99/A024, NUB 3200, EFI-99-16}
\bigskip
\bigskip
\begin{document}
\title{Non-perturbative K3 Orientifolds with NS-NS $B$-flux}
\medskip

\author{Zurab Kakushadze\footnote{E-mail: 
zurab@string.harvard.edu}}

\bigskip
\address{Jefferson Laboratory of Physics, Harvard University,
Cambridge,  MA 02138\\
and\\
Department of Physics, Northeastern University, Boston, MA 02115\\
and\\
Enrico Fermi Institute, University of Chicago, Chicago, IL 60637}
\date{May 5, 1999}
\bigskip
\medskip
\maketitle

\begin{abstract}
{}We consider non-perturbative six dimensional ${\cal N}=1$ space-time
supersymmetric orientifolds of Type IIB on K3 with non-trivial NS-NS $B$-flux.
All of these models are non-perturbative in both orientifold
and heterotic pictures. Thus, some states in such 
compactifications arise in ``twisted'' open string sectors which lack 
world-sheet description in terms of D-branes. We also discuss their 
dual F-theory 
compactifications on certain Voisin-Borcea orbifolds. In particular, the 
explicit construction of non-perturbative K3 orientifolds with NS-NS $B$-flux
gives additional evidence for the conjectured extension of Nikulin's 
classification in the context of Voisin-Borcea orbifolds.
\end{abstract}

{}In the recent years various six (see, {\em e.g.}, \cite{PS,GP,GJ,bij}) and
four (see, {\em e.g.}, \cite{BL,Sagnotti,ZK,KS,Zw,ibanez,KST,223}) dimensional
orientifold vacua have been constructed.  In many cases
the world-sheet approach to orientifolds is adequate and gives rise to
consistent anomaly free vacua in six and four dimensions. 
However, there are cases where the perturbative orientifold 
description misses certain
non-perturbative sectors giving rise to massless states \cite{KST,NP,VB}.
In certain cases this inadequacy results in obvious
inconsistencies such as lack of tadpole and anomaly cancellation. Examples
of such cases were discussed in \cite{Zw,ibanez,KST}. In other cases,
however, the issue is more subtle as the non-perturbative states arise in
anomaly free combinations, so that they are easier to miss. 

{}Recently such non-perturbative orientifolds have been studied in detail 
in \cite{NP,VB}. In particular, six and four dimensional ${\cal N}=1$
supersymmetric non-perturbative 
orientifolds have been constructed in \cite{NP,VB} using 
Type I-heterotic duality and the map between Type IIB orientifolds and 
F-theory as guiding principles. The discussion in \cite{NP,VB} was (mainly)
confined to orientifolds with trivial NS-NS $B$-flux. The purpose of this 
note is to extend the results of \cite{NP,VB} to compactifications with 
non-trivial NS-NS $B$-field turned on. In particular, here we will focus on
six dimensional ${\cal N}=1$ supersymmetric 
non-perturbative orientifolds of Type IIB on K3
with non-trivial NS-NS $B$-flux.  

{}The origin of non-perturbative states in orientifold compactifications
can be understood as follows. Thus, in the K3 orbifold examples 
of \cite{GJ} the orientifold projection is not $\Omega$, which we will use
to denote that in the smooth K3 case, but rather $\Omega J^\prime$, where
$J^\prime$ maps the $g$ twisted sector to its conjugate $g^{-1}$ twisted
sector (assuming $g^2\not=1$) \cite{pol}. Geometrically this can be viewed
as a permutation of two ${\bf P}^1$'s associated with each fixed point of
the orbifold \cite{KST}. (More precisely, these ${\bf P}^1$'s correspond to
the orbifold blow-ups.) This is different from the orientifold projection
in the smooth case where (after blowing up) the orientifold projection
does {\em not} permute the two ${\bf P}^1$'s. In the case of the 
$\Omega J^\prime$ projection the ``twisted'' open string sectors corresponding
to the orientifold elements $\Omega J^\prime g$ are absent
\cite{blum,KST}. However, if the orientifold projection is $\Omega$, then
the ``twisted'' open string sectors corresponding
to the orientifold elements $\Omega g$ are present \cite{KST}. 
In fact, these states
are non-perturbative from the orientifold viewpoint and are required for
gravitational anomaly cancellation in six dimensions. This can be explicitly
seen from the construction of non-perturbative K3 orientifolds without 
the $B$-flux in \cite{NP,VB}.

{}The effect of turning on non-trivial NS-NS $B$-flux in the {\em perturbative}
(that is, ``untwisted'') open string 
sectors is as follows. First, the rank of the
gauge group is reduced by $2^{b/2}$ \cite{Bij}, where $b$ is the rank of the 
matrix $B_{ij}$ ($i,j=1,2,3,4$ label the compact directions corresponding to 
K3). Note that $b$ is always an even integer, and in the case of K3 
compactifications can take values $b=0,2,4$. Second, the multiplicity of the
59 open string sector states 
(if present) is no longer 1 (as in the case without the
$B$-field), but $2^{b/2}$ \cite{bij}. Note that the $B$-field is quantized
in half-integer units (here we are using the normalization where the $B$-field
is defined modulo integer shifts) \cite{Bij,bij}.

{}The effect of the NS-NS $B$-flux in the {\em non-perturbative} (that is,
``twisted'') open string sectors is a bit more non-trivial. The point here
is that Type I-heterotic duality was an important guiding principle in 
determining the twisted open string sector spectrum of non-perturbative K3
orientifolds ({\em e.g.}, in the case of the ${\bf Z}_3$ orbifold limit of K3)
without the $B$-flux
\cite{NP}. In particular, in certain cases the corresponding heterotic dual
is perturbative, which enables one to understand the non-perturbative sectors
on the orientifold side. However, as was pointed out in \cite{NP}, once we
turn on the $B$-field, the above property is no longer there - the heterotic
duals in these cases are no longer perturbative. This makes understanding 
twisted open string sectors in the presence of the $B$-flux more involved.

{}Nonetheless, it turns out that there is a way around this difficulty.    
The key simplification here is due to the observation of \cite{ZK,BN} (also
see \cite{Witten,SS}) that
turning on the $B$-field in an orientifold compactification is equivalent
to turning on non-commuting Wilson lines. Let us be more precise here. For the
sake of definiteness let us consider a compactification on a two-torus $T^2$.
As was explained in detail in \cite{BN}, turning on (quantized) $B$-filed
of rank $b=2$ on $T^2$ is equivalent to modding out by the following freely
acting orbifold. Thus, let us turn on two ${\bf Z}_2$ valued Wilson lines
on $T^2$ - one corresponding to the $a$-cycle of $T^2$, and the other one 
corresponding to the $b$-cycle. Each of these Wilson lines can be described
as a freely acting ${\bf Z}_2$ orbifold which amounts to shifting the $T^2$
lattice by a half-lattice shift along the corresponding cycle of $T^2$. In the
closed string sector the effect of non-zero $B$-field then corresponds to
having non-trivial {\em discrete torsion} between the generators, call them
$g_1$ respectively $g_2$, of the 
two ${\bf Z}_2$ subgroups. In the open string sector each ${\bf Z}_2$ has a 
non-trivial action on the corresponding Chan-Paton charges. 
This action is described by the $16\times 16$ Chan-Paton 
matrices\footnote{Here we work with $16\times 16$ (rather 
than $32\times 32$) Chan-Paton matrices for we choose not to count the 
orientifold images of the corresponding D-branes.} 
$\gamma_{g_1}$ respectively $\gamma_{g_2}$.
The effect of
non-zero $B$-field in the open string sector then corresponds to
having non-commuting Chan-Paton matrices: $\gamma_{g_1}\gamma_{g_2}=-
\gamma_{g_2}\gamma_{g_1}$. That is, even though the (freely acting) 
orbifold group is ${\bf Z}_2\otimes {\bf Z}_2$ (with discrete torsion), the
corresponding embedding in the Chan-Paton charges is in terms of the 
non-Abelian dihedral group $D_4$ (see \cite{BN} for details).

{}It is now clear how to deal with turning on the $B$-field in the twisted 
open string sectors. We start with the cases without the $B$-field, which have
been understood in \cite{NP,VB}, and then mod out by the above freely acting 
orbifold with the corresponding action on the Chan-Paton charges. This 
procedure is relatively straightforward to carry out, albeit there are some 
technical subtleties (such as the fact that various relative signs in the 
corresponding projections require some care) one encounters in the process. 
One also must take into account not only the original twisted sectors modded
out by the above action, but also the additional sectors corresponding to
fixed points of the K3 orbifold group element (which is a pure twist) 
accompanied by the ${\bf Z}_2\otimes {\bf Z}_2$ shifts. For the sake of brevity
here we will skip (most of)
the details of the derivation of the corresponding massless
spectra, and simply summarize the results in various tables. We will, however,
make various clarifying remarks as we discuss the corresponding models.    

{}Thus, consider Type IIB on ${\cal M}_2=T^4/{\bf Z}_N$, where the
generator $g$ of ${\bf Z}_N$, $N=2,3,4,6$, acts on 
the complex coordinates $z_1,z_2$ on 
$T^4$ as follows: 
\begin{equation}
 g z_1=\omega z_1~,~~~g z_2=\omega^{-1} z_2~,
\end{equation}
where $\omega\equiv\exp(2\pi i/N)$. This theory has ${\cal N}=2$
supersymmetry in six dimensions.

{}Next, we consider the $\Omega$ orientifold of this theory. Note that, as
we have already mentioned in the previous section, the $\Omega$ projection 
acts as in the smooth K3 case. This, in particular, implies that we must
first blow up the orbifold singularities before orientifolding.
After orientifolding the closed string sector contains 
the ${\cal N}=1$ supergravity multiplet in six dimensions plus the usual
tensor supermultiplet. We also have some number $n^c_H$ of closed string 
sector 
hypermultiplets plus some number ${\widetilde n}_T$ of extra tensor multiplets.
Note that $n^c_H+{\widetilde n}_T=20$ (this follows from the fact that before
orientifolding in the ${\cal N}=1$ language we have 20 hypermultiplets and
20 tensor multiplets associated with K3). 
The precise values of $n^c_H$ and 
${\widetilde n}_T$, however, depend on the order of the orbifold group $N$ as
well as the rank $b$ of the $B$-field. 

{}Thus, in the ${\bf Z}_2$ case the
untwisted closed string sector gives rise to 4 hypermultiplets only. The ${\bf
Z}_2$ twisted closed string sector produces 16 hypermultiplets and no tensor 
multiplets for $b=0$, 12 hypermultiplets and 4 tensor multiplets for $b=2$, and
10 hypermultiplets and 6 tensor multiplets for $b=4$. This can be seen as 
follows \cite{bij}. Consider a special point in the moduli space where the 
$T^4$ lattice is that of $SO(8)$. In this case the rank of the $B$-field $b=2$.
For this special lattice the ${\bf Z}_2$ twist can be rewritten in terms of
lattice shifts (see \cite{bij} for details), and in the latter language it
is straightforward to see that 12 fixed points are symmetric under the action 
of $\Omega$, whereas 4 of them are antisymmetric. The above conclusion for the
$b=4$ case then follows by considering the four-torus of the factorized form 
$T^2\otimes T^2$ and turning on rank 2 $B$-field in each $T^2$. For each $T^2$
the symmetric and antisymmetric fixed points can be deduced from the $b=2$ case
discussed above as follows. First consider the case where we have $T^2\otimes
T^2$ with rank 2 $B$-field in the first $T^2$ and no $B$-field in the second
$T^2$. Then the second $T^2$ gives 4 fixed points which are symmetric under 
$\Omega$. Since we know that we must have total of 12 symmetric and 4 
antisymmetric fixed points for $b=2$, it then follows that the first $T^2$
gives 3 symmetric and 1 antisymmetric fixed points. Now it is clear that in the
case where both $T^2$'s have rank 2 $B$-field turned on, there are total 10
symmetric and 6 antisymmetric fixed points.

{}Next, let us consider other ${\bf Z}_N$ cases with $N=3,4,6$. 
In the untwisted sector we have 2 hypermultiplets and no tensor multiplets.
For $N=4,6$
we have the ${\bf Z}_2$ twisted sector. As was discussed in \cite{bij}, after
the appropriate projections, in the ${\bf Z}_2$ sector we have: 
10 symmetric and no antisymmetric fixed points for $N=4,b=0$;
8 symmetric
and 2 antisymmetric fixed points for $N=4,b=2$; 7 symmetric and 3 antisymmetric
fixed points for $N=4,b=4$; 6 symmetric and no antisymmetric fixed points for
$N=6$ and all three values of $b=0,2,4$. Finally, in the ${\bf Z}_N$ 
twisted sectors with $N=3,4,6$ the fixed points are always symmetric under
$\Omega$. This can be seen, say, for the ${\bf Z}_3$ case by going to the 
special point in the moduli space where $T^4$ factorizes as $T^2\otimes T^2$
with each $T^2$ corresponding to the $SU(3)$ lattice. In this case each $T^2$
has rank 2 $B$-field turned on. For the $SU(3)$ lattice the ${\bf Z}_3$ twist
can be rewritten in terms of the corresponding ${\bf Z}_3$ valued shift, and 
then it is not difficult to see that all three fixed points in each $T^2$ are
symmetric under $\Omega$.

{}Putting all of the above together, we conclude that for the
non-perturbative $\Omega$ orientifolds of Type IIB on K3 we have:
$n^c_H=20$, ${\widetilde n}_T=0$ for $N=2,b=0$; 
$n^c_H=16$, ${\widetilde n}_T=4$ for $N=2,b=2$;
$n^c_H=14$, ${\widetilde n}_T=6$ for $N=2,b=4$;
$n^c_H=20$, ${\widetilde n}_T=0$ for $N=3,6,b=0,2,4$;
$n^c_H=20$, ${\widetilde n}_T=0$ for $N=4,b=0$;
$n^c_H=18$, ${\widetilde n}_T=2$ for $N=4,b=2$;
$n^c_H=17$, ${\widetilde n}_T=3$ for $N=4,b=4$.
This is reflected in the corresponding tables. 

{}Let us now turn to the open string sector. In the ${\bf Z}_3$ case the 
twisted Chan-Paton matrices are completely fixed by the tadpole cancellation 
conditions for each value of $b$. The corresponding perturbative spectra
can be found in \cite{bij}. As to the non-perturbative ${\bf Z}_3$ twisted 99 
open string sector, in the $b=0$ case it was given in \cite{NP}. To obtain
the corresponding non-perturbative sector states for $b=2,4$, we mod out by the
freely acting orbifold action discussed above. The massless spectra of the 
${\bf Z}_3$ non-perturbative orientifolds with various values of $b$ are 
summarized in Table I.  

{}In the ${\bf Z}_N$ cases with $N=2,4,6$ the tadpole cancellation conditions
do not uniquely fix the twisted Chan-Paton matrices. More precisely, in the
${\bf Z}_2$ case with $b=0$ we have a unique (up to equivalent representations)
solution for $\gamma_R$, where $R$ is the generator of ${\bf Z}_2$:
\begin{equation}
 \gamma_{R}={\mbox{diag}}(i{\bf I}_8,-i{\bf I}_8)~,
\end{equation}
where ${\bf I}_n$ denotes the unit $n\times n$ matrix. (Note that the action
of the orbifold group on the D9- and D5-branes is similar.) However, as
pointed out in \cite{VB}, for $b\not=0$ we have two solutions:
\begin{eqnarray}
 &&\gamma_{R}={\mbox{diag}}(i{\bf I}_{2^{3-b/2}},-i
 {\bf I}_{2^{3-b/2}})~,\\
 &&\gamma_{R}={\mbox{diag}}({\bf I}_{2^{3-b/2}},-
 {\bf I}_{2^{3-b/2}})~,
\end{eqnarray}
where the first solution corresponds to the case without vector structure
\cite{berkooz,intri},
whereas the second solution corresponds to the case with vector structure
\cite{VB}. 

{}In the ${\bf Z}_2$ case there are no non-perturbative twisted open string 
sectors, and the corresponding massless spectra for various values of $b$ can 
be found in \cite{bij} for the cases without vector structure, and in \cite{NP}
for the cases with vector structure. Here we will therefore focus on the
${\bf Z}_4$ and ${\bf Z}_6$ cases with and without vector structure. 
In the cases without vector structure the 
corresponding twisted massless spectra can be worked out by starting from
the $b=0$ case and modding out by the freely acting orbifold action discussed 
above. The resulting massless spectra for the ${\bf Z}_4$ orbifold models
are given in Table I. In Table II we summarize the massless spectra for the 
${\bf Z}_6$ orbifold models without vector structure. Note that all of these
models satisfy the gravitational anomaly cancellation condition 
\cite{anomalies}
\begin{equation}\label{anom}
 n_H-n_V=273-29 n_T~,
\end{equation}
where $n_H$, $n_V$ and $n_T$ are the total numbers of hypermultiplets, vector
multiplets and tensor multiplets, respectively. (Note that $n_T={\widetilde
n}_T+1$.) 

{}Next, we turn to the ${\bf Z}_4$ and ${\bf Z}_6$ cases with vector 
structure. Here we must have $b=2$ or 4. Note that the perturbative open 
string spectrum
of the ${\bf Z}_6$ model with $b=2$ with vector structure is the same as that
of the ${\bf Z}_6$ model with $b=2$ without vector structure \cite{VB}. 
In fact, it is not 
difficult to 
show that the non-perturbative twisted open string spectra are also 
the same in these two cases. The massless spectrum of the ${\bf Z}_6$ case
with $b=4$ with vector structure is summarized in Table III. Finally, the
massless spectra of the ${\bf Z}_4$ models with $b=2$ and $b=4$ with vector
structure are also given in Table III. Note that all of the models in Table III
also satisfy the gravitational anomaly cancellation condition (\ref{anom}).
As to the Abelian anomalies present in models with $U(1)$ factors, they are
expected to be canceled via the generalized Green-Schwarz mechanism
\cite{Sag,berkooz}. 

{}In the remainder of this note we would like to discuss F-theory duals 
of the above non-perturbative orientifolds with NS-NS $B$-flux. In order to
understand these F-theory compactifications, we will take an indirect route.
Let us further compactify a given non-perturbative K3 orientifold on $T^2$.
The resulting four dimensional model has ${\cal N}=2$ supersymmetry. Let us
go to a generic point in the moduli space where the gauge group is maximally 
Higgsed (so that the gauge group is either completely broken or consists of 
Abelian factors only)\footnote{More concretely, first we maximally Higgs
using the hypermultiplet matter, and then use the adjoint scalars in the
${\cal N}=2$ vector multiplets to break any remaining non-Abelian group
to its Cartan subalgebra \cite{KV}.}. 
Let $r(V)$ be the number of open string vector 
multiplets after Higgsing. Then the total number of vector multiplets is given
by $r(V)+n_T+2$, where $n_T$ is the number of tensor multiplets in six 
dimensions. Let $H^0$ be the number of hypermultiplets neutral with respect to
the left-over Abelian gauge group (if any). Then, if we assume that the 
resulting
four dimensional model has a Type IIA dual, the Hodge numbers of the 
corresponding Calabi-Yau three-fold are given by \cite{vafa}:  
\begin{eqnarray}
 &&h^{1,1}=r(V)+n_T+2~,\\
 &&h^{2,1}=H^0-1~.
\end{eqnarray}
Using various dualities between Type IIA, heterotic, Type I and F-theory, it 
is not difficult to see that these Calabi-Yau three-folds must be elliptically 
fibered, and F-theory compactifications on these spaces should be dual to the 
original six dimensional orientifold models. In the table below we give the
Hodge (and Euler) numbers of the Calabi-Yau three-folds for the 
perturbative (that is, $\Omega J^\prime$) K3 orientifolds as well as 
non-perturbative (that is, $\Omega$) K3 orientifolds with various values of
$b$ (note that in the ${\bf Z}_2$ case the $\Omega J^\prime$ and $\Omega$
orientifolds are equivalent as the action of $J^\prime$ is trivial):

\begin{center}
\begin{tabular}{|c|c|c|c|c|} \hline
\phantom{***} Model \phantom{***}
& \phantom{***} $b$ \phantom{***}
& \phantom{***} $(h^{1,1},h^{2,1})$ \phantom{***}
& \phantom{***} $\chi$ \phantom{***} & \phantom{***} $(r,a)$ \phantom{***}\\ 
 \hline
${\bf Z}_2$, $\Omega$ & 0  
            & $(3,243)$ & $-480$ & (2,4)\\ \hline
            & 2  
            & $(7,127)$ & $-240$ & (6,8)\\ \hline
            & 4 
            & $(9,69)$ & $-120$ & (8,10)\\ \hline
${\bf Z}_3$, $\Omega J^\prime$ & 0 
            & $(20,14)$ & $12$ & (11,9)\\ \hline
            & 2 
            & $(16,10)$ & $12$ & (11,11)\\ \hline
            & 4 
            & $(16,10)$ & $12$ & (11,11)\\ \hline
${\bf Z}_4$, $\Omega J^\prime$ & 0 
            & $(7,127)$ & $-240$ & (6,8) \\ \hline
            & 2  
            & $(9,69)$ & $-120$ & (8,10)\\ \hline
            & 4 
            & $(10,40)$ & $-60$ & (9,11)\\ \hline
${\bf Z}_6$, $\Omega J^\prime$ & 0  
            & $(9,69)$ & $-120$ & (6,8)\\ \hline
            & 2 
            & $(9,69)$ & $-120$ & (6,8)\\ \hline
            & 4 
            & $(9,69)$ & $-120$ & (6,8)\\ \hline
${\bf Z}_3$, $\Omega$ & 0 
            & $(11,59)$ & $-96$ & (2,0)\\ \hline
            & 2 
            & $(7,55)$ & $-96$ & (2,2)\\ \hline
            & 4 
            & $(7,55)$ & $-96$ & (2,2)\\ \hline
${\bf Z}_4$, $\Omega$ & 0 
            & $(3,243)$ & $-480$ & (2,4) \\ \hline
            & 2  
            & $(5,185)$ & $-360$ & (4,6)\\ \hline
            & 4 
            & $(6,156)$ & $-300$ & (5,7)\\ \hline
${\bf Z}_6$, $\Omega$ & 0  
            & $(3,243)$ & $-480$ & (2,4)\\ \hline
            & 2 
            & $(3,243)$ & $-480$ & (2,4)\\ \hline
            & 4 
            & $(3,243)$ & $-480$ & (2,4)\\ \hline
\end{tabular}
\end{center}
Note that the above table covers the cases with and without vector structure
alike. In the fourth column we have displayed the Euler number 
$\chi=2(h^{1,1}-h^{2,1})$. The meaning of the last column will become clear
in a moment. 

{}All of the above Calabi-Yau three-folds correspond to Voisin-Borcea orbifolds
\cite{Voisin,Borcea} (for a physicists discussion, see, {\em e.g.}, 
\cite{MV,Asp}). Here we will therefore briefly review some facts about 
Voisin-Borcea orbifolds.

{}Thus, let ${\cal W}_2$ be a K3 surface (which is not necessarily an 
orbifold of $T^4$) which admits an involution $J$ such that it reverses the 
sign of the holomorphic
two-form $\Omega_2$ on ${\cal W}_2$. Consider the following quotient:
\begin{equation}
 {\cal Y}_3= (T^2\otimes {\cal W}_2)/Y~,
\end{equation}
where $Y=\{1,S\}\approx {\bf Z}_2$, and $S$ acts as $Sz_0=-z_0$ on $T^2$ 
($z_0$ being a complex coordinate on $T^2$), and as $J$ on ${\cal W}_2$. This
quotient, known as a Voisin-Borcea orbifold, 
is a Calabi-Yau three-fold with $SU(3)$ holonomy 
which is elliptically 
fibered over the base ${\cal B}_2={\cal W}_2/B$, where 
$B=\{1,J\}\approx{\bf Z}_2$. 

{}Nikulin gave a classification \cite{Nik} 
of possible involutions of K3 surfaces in terms of three 
invariants $(r,a,\delta)$.
The result of this classification is plotted in Fig.1 (which has been borrowed
from \cite{KST})
according to the values of $r$ 
and $a$. The open and closed circles correspond to the cases with $\delta=0$
and $\delta=1$, respectively. (The cases denoted by ``$\otimes$'' are outside
of Nikulin's classification, and we will discuss them shortly.) In the case 
$(r,a,\delta)=(10,10,0)$ the base ${\cal B}_2$ is an Enriques surface, and the 
corresponding ${\cal Y}_3$ has Hodge numbers $(h^{1,1},h^{2,1})=(11,11)$.
In all the other cases the Hodge numbers are given by:
\begin{eqnarray}\label{hodge1}
 &&h^{1,1}=5+3r-2a~,\\
 \label{hodge2}
 &&h^{2,1}=65-3r-2a~.
\end{eqnarray} 

{}For $(r,a,\delta)=(10,10,0)$ the ${\bf Z}_2$ twist $S$ is freely acting 
(that is, it has no fixed points). 
For $(r,a,\delta)=(10,8,0)$ the fixed point set
of $S$ consists of two curves of genus 1. The base ${\cal B}_2$ in this case 
is ${\bf P}^2$ blown up at 9 points. In all the other cases the fixed point set
of $S$ consists of one curve of genus $g$ plus $k$ rational curves, where
\begin{eqnarray}
 &&g={1\over 2}(22-r-a)~,\\
 &&k={1\over 2}(r-a)~.
\end{eqnarray}

{}Note that except for the cases with $a=22-r$, 
$r=11,\dots,20$, the mirror pair
of ${\cal Y}_3$ is given by the Voisin-Borcea orbifold 
${\widetilde {\cal Y}}_3$ with ${\widetilde r}=20-r$, ${\widetilde a}=a$.
Under the mirror transform we have: ${\widetilde g}=f$, ${\widetilde f}=g$, 
where $f\equiv k+1$. 

{}In the cases $a=22-r$, $r=11,\dots,20$, the mirror would have to have
${\widetilde r}=20-r$ and ${\widetilde a}=a={\widetilde r}+2$, where 
${\widetilde r}=0,\dots,9$. We have depicted these cases in Fig.2 using 
the ``$\otimes$'' symbol. In 
particular, 
we have plotted cases with $a=r+2$, $r=0,\dots,10$.
The Hodge numbers for these 
cases are still given by (\ref{hodge1}) and (\ref{hodge2}) 
(which follows from
their definition as mirror pairs of the cases with $a=22-r$, 
$r=11,\dots,20$). (This is
true for $a=r+2$, $r=0,\dots,9$. 
Extrapolation to $r=10$ is motivated by 
the fact that
in this case we get $(h^{1,1},h^{2,1})=(11,11)$ which is the same as for 
$(r,a,\delta)=(10,10,0)$.) In \cite{KST} it was argued that these Voisin-Borcea
orbifolds also exist, albeit they are {\em singular}. 
In fact, some of them can be constructed explicitly (see
\cite{KST} for details). For instance, the case with $r=2,a=4$ has Hodge 
numbers $(h^{1,1},h^{2,1})=(3,51)$, which correspond to the singular
Calabi-Yau realized as the ${\bf Z}_2\otimes{\bf Z}_2$ orbifold with discrete
torsion \cite{FIQ,VW}.

{}As we have already mentioned, here we would like to briefly review
F-theory compactifications on Voisin-Borcea orbifolds (which correspond to
${\cal N}=1$ supersymmetric vacua in six dimensions). Thus, consider F-theory
on ${\cal Y}_3$ with $(r,a,\delta)\not=
(10,10,0)$ or $(10,8,0)$. This
gives rise to the following massless spectrum in six 
dimensions. The number of tensor multiplets is $T=r-1$. The number of neutral 
hypermultiplets is $H=22-r$. The gauge group is $SO(8)\otimes SO(8)^k$.
There are $g$ adjoint hypermultiplets of the first $SO(8)$. There are no 
hypermultiplets charged under the other $k$ $SO(8)$'s. Under mirror
symmetry\footnote{For a discussion of F-theory compactifications on mirror
manifolds, see, {\em e.g.}, \cite{Perevalov}.} 
$g$ and $f=k+1$ are interchanged. Thus, the vector multiplets in the adjoint
of $SO(8)^k$ are traded for $g-1$ hypermultiplets in the adjoint of the first 
$SO(8)$. That is, gauge symmetry turns 
into global symmetry and {\em vice-versa}.
This can be pushed further to understand F-theory compactifications on
Calabi-Yau three-folds 
with $a=r+2$, $r=1,\dots,10$, which give the following 
spectra.
The number of tensor multiplets is $T=r-1$. There are 
$H=22-r$ neutral hypermultiplets. In 
addition there are $g=10-r$ hypermultiplets
transforming as adjoints under a global $SO(8)$ symmetry. 
There are no gauge bosons,
however. It is not difficult to verify that this massless 
spectrum is free of gravitational
anomalies in six dimensions. In fact, in \cite{KST} it was 
conjectured that all of these singular Calabi-Yau three-folds exist, and
F-theory compactifications
on these singular spaces are equivalent to F-theory compactifications
on smooth Calabi-Yau three-folds according to the following relation 
\cite{KST}:
\begin{eqnarray}
 &&{\mbox{F-theory on 
 ${\cal Y}_3$ with $(h^{1,1},h^{2,1})=(r+1,61-5r)$ is equivalent
 to}}\nonumber\\
 &&{\mbox{F-theory on 
 ${\widehat{\cal Y}}_3$ with $({\hat h}^{1,1},{\hat h}^{2,1})
 =(r+1,301-29r)$  ($r=1,\dots,9$)}}~.
\end{eqnarray}
Thus, for instance, 
for $r=2$ we get $({\hat h}^{1,1},{\hat h}^{2,1})=(3,243)$, 
which is the elliptic Calabi-Yau given by the $T^2$ fibration over
the base ${\bf P}^1\otimes {\bf P}^1$. 

{}In the above discussion of {\em perturbative} K3 orientifolds
we have encountered Calabi-Yau three-folds
with the Hodge numbers $({\hat h}^{1,1},{\hat h}^{2,1})=
(3,243),(7,127),(9,69),(10,40)$, 
and also $(h^{1,1},h^{2,1})=(20,14),(16,10)$ (see the above table). 
The last two cases are within
Nikulin's classification, and correspond to $(r,a)=(11,9)$ and $(11,11)$,
respectively. The first four cases, however, are outside of Nikulin's 
classification, and correspond to the above Calabi-Yau three-folds 
${\widehat{\cal Y}}_3$ (whose Hodge numbers are given by
$({\hat h}^{1,1},{\hat h}^{2,1})=(r+1,301-r)$) with $r=2,6,8,9$, respectively.
The cases $r=6,8$ were originally discussed in \cite{GJ1}. In fact, in 
\cite{BG} the Calabi-Yau three-fold with $({\hat h}^{1,1},{\hat h}^{2,1})=
(7,127)$ (corresponding to $r=6$) was explicitly constructed. Here we also 
have an additional case with $r=9$. The fact that perturbative K3 orientifolds
with NS-NS $B$-field are consistent string backgrounds gives 
evidence for the conjectured extension of Nikulin's classification.

{}In fact, in our discussion of {\em non-perturbative} K3 orientifolds 
we have encountered four new cases: $({\hat h}^{1,1},{\hat h}^{2,1})=(5,185),
(6,156)$, and $(h^{1,1},h^{2,1})=(11,59),(7,55)$. The first two cases 
correspond to the ${\widehat {\cal Y}}_3$ three-folds with $r=4,5$, 
respectively. Our construction of non-perturbative K3 orientifolds with the
NS-NS $B$-flux therefore gives additional evidence for the above conjecture.
As to the other two cases, they correspond to Voisin-Borcea orbifolds (within
Nikulin's classification) with $(r,a)=(2,0)$ (in this case the base is the
Hirzebruch surface ${\bf F}_4$) respectively $(r,a)=(2,2)$ (in which case
the base is either ${\bf F}_0$ or ${\bf F}_1$).  

{}Before we end this note, we would like to point out that, having understood
non-perturbative K3 orientifolds with NS-NS $B$-flux, it is 
now straightforward to construct non-perturbative orientifolds on $T^6/\Gamma$
orbifolds with $SU(3)$ holonomy in the presence of NS-NS $B$-flux along the 
lines of \cite{NP,VB}.

{}This work was supported in part by the grant
NSF PHY-96-02074, 
and the DOE 1994 OJI award. I would also like to thank Albert and 
Ribena Yu for financial support.

%%%%%%%%%%%%%Table I %%%%%%%%
%%%%%%%%%%%%%%%%%%%%%%%%%%%%%%%%%%%%%%%%%%%%%%%%%%%%%%%%%%%%%%%%%%%%%%%%%%%%%%%
\begin{table}[t]
\begin{tabular}{|c|c|c|l|c|c|}
%%%%%%%%%%%%%%%%%%%%%%%%%%%%%%%%%%%%%%%%%%%%%%%%%%%%%%%%%%%%%%%%%%%%%%%%%%%%
Model & $b$ & Gauge Group & \phantom{Hy} Charged  & Neutral 
& Extra Tensor  \\
  &   &             &Hypermultiplets & Hypermultiplets
&Multiplets \\
\hline
%%%%%%%%%%%%%%%%%%%%%%%%%%%%%%%%%%%%%%%%%%%%%%%%%%%%%%%%%%%%%%%%%%%%%%%%%%%
${\bf Z}_3$ & 0 & $[U(8) \otimes SO(16)]_{99}$ & $({\bf 28},{\bf 1})_U$ & $20$
& $0$ \\
& && $({\bf 8},{\bf 16})_U$ & & \\
& && $9\times ({\bf 28},{\bf 1})_T$ & & \\
\hline
%%%%%%%%%%%%%%%%%%%%%%%%%%%%%%%%%%%%%%%%%%%%%%%%%%%%%%%%%%%%%%%%%%%%%%%%%%%
& 2 & $U(8)_{99}$ & ${\bf 36}_U$ & $20$
& $0$ \\
& & & $9\times {\bf 28}_T$ & & \\
\hline
%%%%%%%%%%%%%%%%%%%%%%%%%%%%%%%%%%%%%%%%%%%%%%%%%%%%%%%%%%%%%%%%%%%%%%%%%%%
& 4 & $SO(8)_{99}$ & $9\times {\bf 28}_T$ & $20$
& $0$ \\
\hline
%%%%%%%%%%%%%%%%%%%%%%%%%%%%%%%%%%%%%%%%%%%%%%%%%%%%%%%%%%%%%%%%%%%%%%%%
${\bf Z}_4$ &0 &$[U(8) \otimes U(8)]_{99}\otimes$ 
& $({\bf 28},{\bf 1};{\bf 1},{\bf 1})_U$ & $20$ & $0$ \\
& &$[U(8) \otimes U(8)]_{55}$&
$({\bf 1},{\bf 28};{\bf 1},{\bf 1})_U$  & & \\
& && $({\bf 8},{\bf 8};{\bf 1},{\bf 1})_U$ & & \\
& && $({\bf 28},{\bf 1};{\bf 1},{\bf 1})_T$ & & \\
& && $({\bf 1},{\bf 28};{\bf 1},{\bf 1})_T$ & & \\
& && same as above with $99\leftrightarrow 55$ & & \\
& && $({\bf 8},{\bf 1};{\bf 8},{\bf 1})_U$ & & \\
& && $({\bf 1},{\bf 8};{\bf 1},{\bf 8})_U$ & & \\
\hline
%%%%%%%%%%%%%%%%%%%%%%%%%%%%%%%%%%%%%%%%%%%%%%%%%%%%%%%%%%%%%%%%%%%%%%%%
&2 &$[U(4) \otimes U(4)]_{99}\otimes$ 
& $({\bf 6},{\bf 1};{\bf 1},{\bf 1})_U$ & $18$ & $2$ \\
& &$[U(4) \otimes U(4)]_{55}$&
$({\bf 1},{\bf 6};{\bf 1},{\bf 1})_U$  & & \\
& && $({\bf 4},{\bf 4};{\bf 1},{\bf 1})_U$ & & \\
& && $({\bf 10},{\bf 1};{\bf 1},{\bf 1})_T$ & & \\
& && $3\times ({\bf 6},{\bf 1};{\bf 1},{\bf 1})_T$ & & \\
& && $({\bf 1},{\bf 10};{\bf 1},{\bf 1})_T$ & & \\
& && $3\times ({\bf 1},{\bf 6};{\bf 1},{\bf 1})_T$ & & \\
& && same as above with $99\leftrightarrow 55$ & & \\
& && $2\times ({\bf 4},{\bf 1};{\bf 4},{\bf 1})_U$ & & \\
& && $2\times ({\bf 1},{\bf 4};{\bf 1},{\bf 4})_U$ & & \\
\hline
%%%%%%%%%%%%%%%%%%%%%%%%%%%%%%%%%%%%%%%%%%%%%%%%%%%%%%%%%%%%%%%%%%%%%%%%
&4 &$[U(2) \otimes U(2)]_{99}\otimes$ 
& $4\times ({\bf 1},{\bf 1};{\bf 1},{\bf 1})_U$ & $17$ & $3$ \\
& &$[U(2) \otimes U(2)]_{55}$&
$({\bf 2},{\bf 2};{\bf 1},{\bf 1})_U$  & & \\
& && $6\times ({\bf 3},{\bf 1};{\bf 1},{\bf 1})_T$ & & \\
& && $10\times ({\bf 1},{\bf 1};{\bf 1},{\bf 1})_T$ & & \\
& && $6\times ({\bf 1},{\bf 3};{\bf 1},{\bf 1})_T$ & & \\
& && $10\times ({\bf 1},{\bf 1};{\bf 1},{\bf 1})_T$ & & \\
& && same as above with $99\leftrightarrow 55$ & & \\
& && $4\times ({\bf 2},{\bf 1};{\bf 2},{\bf 1})_U$ & & \\
& && $4\times ({\bf 1},{\bf 2};{\bf 1},{\bf 2})_U$ & & \\
\hline
%%%%%%%%%%%%%%%%%%%%%%%%%%%%%%%%%%%%%%%%%%%%%%%%%%%%%%%%%%%%%%%%%%
\end{tabular}
%%%%%%%%%%%%%%%%%%%%%%%%%%%%%%%%%%%%%%%%%%%%%%%%%%%%%%%%%%%%%%%%%%%%%%%%
%%%%%%%%%%%%%%%%%%%%%%%%%%%%%%%%%%%%%%%%%%%%%%%%%%%%%%%%%%%%%%%%%%%%%%%%%%%
\caption{The massless spectra of the non-perturbative Type IIB orientifolds
on $T^4/{\bf Z}_N$ for $N=3,4$ with the NS-NS $B$-field of rank $b$.
The semi-colon in the column ``Charged Hypermultiplets'' separates $99$ and 
$55$ representations. The subscript ``$U$'' indicates that the
corresponding (``untwisted'') state is perturbative from the orientifold
viewpoint. The subscript ``$T$'' indicates that the
corresponding (``twisted'') state is non-perturbative from the orientifold
viewpoint. The $U(1)$ charges are not shown, and by ``neutral''
hypermultiplets we mean that the corresponding states are not charged
under the non-Abelian subgroups.}
\end{table}
%%%%%%%%%%%%%%%%%%%%%%%%%%%%%%%%%%%%%%%%%%%%%%%%%%%%%%%%%%%%%%%%%%%%%%%%%%%%%%%

%%%%%%%%%%%%%Table II %%%%%%%%
%%%%%%%%%%%%%%%%%%%%%%%%%%%%%%%%%%%%%%%%%%%%%%%%%%%%%%%%%%%%%%%%%%%%%%%%%%%%%%%
\begin{table}[t]
\begin{tabular}{|c|c|c|l|c|c|}
%%%%%%%%%%%%%%%%%%%%%%%%%%%%%%%%%%%%%%%%%%%%%%%%%%%%%%%%%%%%%%%%%%%%%%%%%%%%
Model & $b$ & Gauge Group & \phantom{Hy} Charged  & Neutral 
& Extra Tensor  \\
  &   &             &Hypermultiplets & Hypermultiplets
&Multiplets \\
\hline
%%%%%%%%%%%%%%%%%%%%%%%%%%%%%%%%%%%%%%%%%%%%%%%%%%%%%%%%%%%%%%%%%%%%%%%%
${\bf Z}_6$ &0 & $[U(4) \otimes U(4) \otimes U(8)]_{99}\otimes$ 
& $({\bf 6},{\bf 1},{\bf 1};{\bf 1},{\bf 1},{\bf 1})_U$ & $20$ & $0$ \\
&& $[U(4) \otimes U(4) \otimes U(8)]_{55}$&
$({\bf 1},{\bf 6},{\bf 1};{\bf 1},{\bf 1},{\bf 1})_U$  & & \\
& && $({\bf 4},{\bf 1},{\bf 8};{\bf 1},{\bf 1},{\bf 1})_U$ & & \\
& && $({\bf 1},{\bf 4},{\bf 8};{\bf 1},{\bf 1},{\bf 1})_U$ & & \\
& && same as above with $99\leftrightarrow 55$ & &\\
& && $({\bf 4},{\bf 1},{\bf 1};{\bf 4},{\bf 1},{\bf 1})_U$ & & \\
& && $({\bf 1},{\bf 4},{\bf 1};{\bf 1},{\bf 4},{\bf 1})_U$ & & \\
& && $({\bf 1},{\bf 1},{\bf 8};{\bf 1},{\bf 1},{\bf 8})_U$ & & \\
& && $5\times ({\bf 6},{\bf 1},{\bf 1};{\bf 1},{\bf 1},{\bf 1})_T$ & & \\
& && $5\times ({\bf 1},{\bf 6},{\bf 1};{\bf 1},{\bf 1},{\bf 1})_T$ & & \\
& && $4\times ({\bf 4},{\bf 4},{\bf 1};{\bf 1},{\bf 1},{\bf 1})_T$ & & \\
& && $({\bf 1},{\bf 1},{\bf 1};{\bf 6},{\bf 1},{\bf 1})_T$ & & \\
& && $({\bf 1},{\bf 1},{\bf 1};{\bf 1},{\bf 6},{\bf 1})_T$ & & \\
& && $({\bf 4},{\bf 1},{\bf 1};{\bf 4},{\bf 1},{\bf 1})_T$ & & \\
& && $({\bf 1},{\bf 4},{\bf 1};{\bf 1},{\bf 4},{\bf 1})_T$ & & \\
\hline
%%%%%%%%%%%%%%%%%%%%%%%%%%%%%%%%%%%%%%%%%%%%%%%%%%%%%%%%%%%%%%%%%%%%%%%%
%%%%%%%%%%%%%%%%%%%%%%%%%%%%%%%%%%%%%%%%%%%%%%%%%%%%%%%%%%%%%%%%%%%%%%%%
 &2 & $[U(4) \otimes U(4)]_{99}\otimes$ 
& $({\bf 6},{\bf 1};{\bf 1},{\bf 1})_U$ & $20$ & $0$ \\
&& $[U(4) \otimes U(4)]_{55}$&
$({\bf 1},{\bf 6};{\bf 1},{\bf 1})_U$  & & \\
& && $({\bf 4},{\bf 4};{\bf 1},{\bf 1})_U$ & & \\
& && same as above with $99\leftrightarrow 55$ & &\\
& && $2\times ({\bf 4},{\bf 1};{\bf 4},{\bf 1})_U$ & & \\
& && $2\times ({\bf 1},{\bf 4};{\bf 1},{\bf 4})_U$ & & \\
& && $5\times ({\bf 6},{\bf 1};{\bf 1},{\bf 1})_T$ & & \\
& && $5\times ({\bf 1},{\bf 6};{\bf 1},{\bf 1})_T$ & & \\
& && $4\times ({\bf 4},{\bf 4};{\bf 1},{\bf 1})_T$ & & \\
& && $({\bf 1},{\bf 1};{\bf 6},{\bf 1})_T$ & & \\
& && $({\bf 1},{\bf 1};{\bf 1},{\bf 6})_T$ & & \\
& && $({\bf 4},{\bf 1};{\bf 4},{\bf 1})_T$ & & \\
& && $({\bf 1},{\bf 4};{\bf 1},{\bf 4})_T$ & & \\
\hline
%%%%%%%%%%%%%%%%%%%%%%%%%%%%%%%%%%%%%%%%%%%%%%%%%%%%%%%%%%%%%%%%%%%%
 &4 & $U(4)_{99}\otimes U(4)_{55}$ 
& $2\times ({\bf 6};{\bf 1})_U$ & $20$ & $0$ \\
&& &
$2\times ({\bf 1};{\bf 6})_U$  & & \\
& && $4\times ({\bf 4};{\bf 4})_U$ & & \\
& && $10\times ({\bf 6};{\bf 1})_T$ & & \\
& && $4\times ({\bf 16};{\bf 1})_T$ & & \\
& && $2\times ({\bf 1};{\bf 6})_T$ & & \\
& && $2\times ({\bf 4};{\bf 4})_T$ & & \\
\hline
\end{tabular}
%%%%%%%%%%%%%%%%%%%%%%%%%%%%%%%%%%%%%%%%%%%%%%%%%%%%%%%%%%%%%%%%%%%%%%%%%%%
\caption{The massless spectra of the non-perturbative
Type IIB orientifolds
on $T^4/{\bf Z}_6$ with the NS-NS $B$ field of rank $b$.
The semi-colon in the column ``Charged Hypermultiplets'' separates $99$ and 
$55$ representations. The subscript ``$U$'' indicates that the
corresponding (``untwisted'') state is perturbative from the orientifold
viewpoint. The subscript ``$T$'' indicates that the
corresponding (``twisted'') state is non-perturbative from the orientifold
viewpoint. The $U(1)$ charges are not shown, and by ``neutral''
hypermultiplets we mean that the corresponding states are not charged
under the non-Abelian subgroups. Note that ${\bf 16}$ is a {\em reducible}
representation of $SU(4)$.}
\end{table}
%%%%%%%%%%%%%%%%%%%%%%%%%%%%%%%%%%%%%%%%%%%%%%%%%%%%%%%%%%%%%%%%%%%%%%%%%%%%%%%

%%%%%%%%%%%%%Table III %%%%%%%%
%%%%%%%%%%%%%%%%%%%%%%%%%%%%%%%%%%%%%%%%%%%%%%%%%%%%%%%%%%%%%%%%%%%%%%%%%%%%%%%
\begin{table}[t]
\begin{tabular}{|c|c|c|l|c|c|}
%%%%%%%%%%%%%%%%%%%%%%%%%%%%%%%%%%%%%%%%%%%%%%%%%%%%%%%%%%%%%%%%%%%%%%%%%%%%
Model & $b$ & Gauge Group & \phantom{Hy} Charged  & Neutral 
& Extra Tensor  \\
  &   &             &Hypermultiplets & Hypermultiplets
&Multiplets \\
\hline
%%%%%%%%%%%%%%%%%%%%%%%%%%%%%%%%%%%%%%%%%%%%%%%%%%%%%%%%%%%%%%%%%%%%%%%%
${\bf Z}_4$, VS &2 & $[U(4) \otimes Sp(4) \otimes Sp(4)]_{99}\otimes$ 
& $({\bf 4},{\bf 4},{\bf 1};{\bf 1},{\bf 1},{\bf 1})_U$ & $18$ & $2$ \\
&& $[U(4) \otimes Sp(4) \otimes Sp(4)]_{55}$&
$({\bf 4},{\bf 1},{\bf 4};{\bf 1},{\bf 1},{\bf 1})_U$  & & \\
& && $({\bf 4},{\bf 1},{\bf 1};{\bf 1},{\bf 4},{\bf 1})_U$ & & \\
& && $({\bf 4},{\bf 1},{\bf 1};{\bf 1},{\bf 1},{\bf 4})_U$ & & \\
& && $2\times ({\bf 6},{\bf 1},{\bf 1};{\bf 1},{\bf 1},{\bf 1})_T$ & & \\
& && $({\bf 1},{\bf 6},{\bf 1};{\bf 1},{\bf 1},{\bf 1})_T$ & & \\
& && $({\bf 1},{\bf 1},{\bf 6};{\bf 1},{\bf 1},{\bf 1})_T$ & & \\
& && $({\bf 4},{\bf 4},{\bf 1};{\bf 1},{\bf 1},{\bf 1})_T$ & & \\
& && $({\bf 4},{\bf 1},{\bf 4};{\bf 1},{\bf 1},{\bf 1})_T$ & & \\
& && same as above with $99\leftrightarrow 55$ & &\\
\hline
%%%%%%%%%%%%%%%%%%%%%%%%%%%%%%%%%%%%%%%%%%%%%%%%%%%%%%%%%%%%%%%%%%%%%%%%
${\bf Z}_4$, VS &4 & $[U(2) \otimes Sp(2) \otimes Sp(2)]_{99}\otimes$ 
& $({\bf 2},{\bf 2},{\bf 1};{\bf 1},{\bf 1},{\bf 1})_U$ & $17$ & $3$ \\
&& $[U(2) \otimes Sp(2) \otimes Sp(2)]_{55}$&
$({\bf 2},{\bf 1},{\bf 2};{\bf 1},{\bf 1},{\bf 1})_U$  & & \\
& && $2\times ({\bf 2},{\bf 1},{\bf 1};{\bf 1},{\bf 2},{\bf 1})_U$ & & \\
& && $2\times ({\bf 2},{\bf 1},{\bf 1};{\bf 1},{\bf 1},{\bf 2})_U$ & & \\
& && $2\times ({\bf 3},{\bf 1},{\bf 1};{\bf 1},{\bf 1},{\bf 1})_T$ & & \\
& && $({\bf 1},{\bf 3},{\bf 1};{\bf 1},{\bf 1},{\bf 1})_T$ & & \\
& && $({\bf 1},{\bf 1},{\bf 3};{\bf 1},{\bf 1},{\bf 1})_T$ & & \\
& && $4\times ({\bf 2},{\bf 2},{\bf 1};{\bf 1},{\bf 1},{\bf 1})_T$ & & \\
& && $4\times ({\bf 2},{\bf 1},{\bf 2};{\bf 1},{\bf 1},{\bf 1})_T$ & & \\
& && same as above with $99\leftrightarrow 55$ & &\\
& && $24\times ({\bf 1},{\bf 1},{\bf 1};{\bf 1},{\bf 1},{\bf 1})_T$ & & \\
\hline
%%%%%%%%%%%%%%%%%%%%%%%%%%%%%%%%%%%%%%%%%%%%%%%%%%%%%%%%%%%%%%%%%%%%%%%%
%%%%%%%%%%%%%%%%%%%%%%%%%%%%%%%%%%%%%%%%%%%%%%%%%%%%%%%%%%%%%%%%%%%%%%%%
${\bf Z}_6$, VS &4 & $[Sp(4) \otimes Sp(4)]_{99}\otimes$ 
& $({\bf 4},{\bf 4};{\bf 1},{\bf 1})_U$ & $20$ & $0$ \\
&& $[Sp(4) \otimes Sp(4)]_{55}$&
$({\bf 1},{\bf 1};{\bf 4},{\bf 4})_U$  & & \\
& && $2\times ({\bf 4},{\bf 1};{\bf 4},{\bf 1})_U$ & & \\
& && $2\times ({\bf 1},{\bf 4};{\bf 1},{\bf 4})_U$ & & \\
& && $5\times ({\bf 6},{\bf 1};{\bf 1},{\bf 1})_T$ & & \\
& && $5\times ({\bf 1},{\bf 6};{\bf 1},{\bf 1})_T$ & & \\
& && $4\times ({\bf 4},{\bf 4};{\bf 1},{\bf 1})_T$ & & \\
& && $({\bf 1},{\bf 1};{\bf 6},{\bf 1})_T$ & & \\
& && $({\bf 1},{\bf 1};{\bf 1},{\bf 6})_T$ & & \\
& && $({\bf 4},{\bf 1};{\bf 4},{\bf 1})_T$ & & \\
& && $({\bf 1},{\bf 4};{\bf 1},{\bf 4})_T$ & & \\
\hline
%%%%%%%%%%%%%%%%%%%%%%%%%%%%%%%%%%%%%%%%%%%%%%%%%%%%%%%%%%%%%%%%%%%%
\end{tabular}
%%%%%%%%%%%%%%%%%%%%%%%%%%%%%%%%%%%%%%%%%%%%%%%%%%%%%%%%%%%%%%%%%%%%%%%%%%%
\caption{The massless spectra of the non-perturbative
Type IIB orientifolds
on $T^4/{\bf Z}_N$ for $N=4,6$ with the NS-NS $B$ field of rank $b$.
The semi-colon in the column ``Charged Hypermultiplets'' separates $99$ and 
$55$ representations. The subscript ``$U$'' indicates that the
corresponding (``untwisted'') state is perturbative from the orientifold
viewpoint. The subscript ``$T$'' indicates that the
corresponding (``twisted'') state is non-perturbative from the orientifold
viewpoint. The $U(1)$ charges are not shown, and by ``neutral''
hypermultiplets we mean that the corresponding states are not charged
under the non-Abelian subgroups. Note that ${\bf 6}$ is a {\em reducible} representation of $Sp(4)$ (in our conventions the rank of $Sp(2n)$ is $n$).}
\end{table}
%%%%%%%%%%%%%%%%%%%%%%%%%%%%%%%%%%%%%%%%%%%%%%%%%%%%%%%%%%%%%%%%%%%%%%%%%%%%%%%

%%%%%%%%%%%%%%%% FIGURE 1 %%%%%%%%%%%%%%%
\newpage
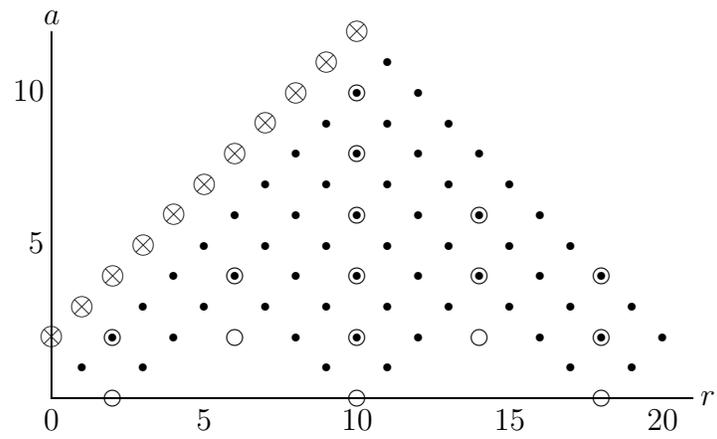
\begin{figure}
\setlength{\unitlength}{0.008in}%
$$\begin{picture}(445,266)(60,385)
\thinlines
\put(72,435){$\otimes$}
\put(92,455){$\otimes$}
\put(112,475){$\otimes$}
\put(132,495){$\otimes$}
\put(152,515){$\otimes$}
\put(172,535){$\otimes$}
\put(192,555){$\otimes$}
\put(212,575){$\otimes$}
\put(232,595){$\otimes$}
\put(252,615){$\otimes$}
\put(272,635){$\otimes$}

\put(100,420){\circle*{6}}

\put(140,420){\circle*{6}}
\put(260,420){\circle*{6}}
\put(300,420){\circle*{6}}
\put(420,420){\circle*{6}}
\put(460,420){\circle*{6}}
\put(120,440){\circle*{6}}
\put(160,440){\circle*{6}}
\put(240,440){\circle*{6}}
\put(280,440){\circle*{6}}
\put(320,440){\circle*{6}}
\put(400,440){\circle*{6}}
\put(440,440){\circle*{6}}
\put(140,460){\circle*{6}}
\put(180,460){\circle*{6}}
\put(220,460){\circle*{6}}
\put(260,460){\circle*{6}}
\put(300,460){\circle*{6}}
\put(340,460){\circle*{6}}
\put(380,460){\circle*{6}}
\put(420,460){\circle*{6}}
\put(160,480){\circle*{6}}
\put(200,480){\circle*{6}}
\put(240,480){\circle*{6}}
\put(280,480){\circle*{6}}
\put(360,480){\circle*{6}}
\put(400,480){\circle*{6}}
\put(180,500){\circle*{6}}
\put(220,500){\circle*{6}}
\put(260,500){\circle*{6}}
\put(300,500){\circle*{6}}
\put(340,500){\circle*{6}}
\put(380,500){\circle*{6}}
\put(200,520){\circle*{6}}
\put(240,520){\circle*{6}}
\put(280,520){\circle*{6}}
\put(320,520){\circle*{6}}
\put(360,520){\circle*{6}}
\put(220,540){\circle*{6}}
\put(260,540){\circle*{6}}
\put(300,540){\circle*{6}}
\put(340,540){\circle*{6}}
\put(240,560){\circle*{6}}
\put(280,560){\circle*{6}}
\put(320,560){\circle*{6}}
\put(260,580){\circle*{6}}
\put(300,580){\circle*{6}}
\put(280,600){\circle*{6}}
\put(320,480){\circle*{6}}
\put(120,400){\circle{10}}
\put(280,400){\circle{10}}
\put(440,400){\circle{10}}
\put(200,440){\circle{10}}
\put(200,480){\circle{10}}
\put(360,440){\circle{10}}
\put(360,480){\circle{10}}
\put(280,480){\circle{10}}
\put(280,520){\circle{10}}
\put(280,560){\circle{10}}
\put(280,600){\circle{10}}
\put(280,440){\circle{10}}
\put(120,440){\circle{10}}
\put(440,440){\circle{10}}
\put( 80,400){\line( 1, 0){420}}
\put( 80,400){\line( 0, 1){240}}
\put(300,620){\circle*{6}}
\put(320,600){\circle*{6}}
\put(340,580){\circle*{6}}
\put(380,540){\circle*{6}}
\put(400,520){\circle*{6}}
\put(420,500){\circle*{6}}
\put(440,480){\circle*{6}}
\put(440,480){\circle{10}}
\put(460,460){\circle*{6}}
\put(480,440){\circle*{6}}
\put(360,520){\circle{10}}
\put(360,560){\circle*{6}}
\put( 75,379){\makebox(0,0)[lb]{\raisebox{0pt}[0pt][0pt]{0}}}
\put(175,379){\makebox(0,0)[lb]{\raisebox{0pt}[0pt][0pt]{5}}}
\put(270,379){\makebox(0,0)[lb]{\raisebox{0pt}[0pt][0pt]{10}}}
\put(370,379){\makebox(0,0)[lb]{\raisebox{0pt}[0pt][0pt]{15}}}
\put(470,379){\makebox(0,0)[lb]{\raisebox{0pt}[0pt][0pt]{20}}}
\put( 65,495){\makebox(0,0)[lb]{\raisebox{0pt}[0pt][0pt]{5}}}
\put( 55,595){\makebox(0,0)[lb]{\raisebox{0pt}[0pt][0pt]{10}}}
\put(505,395){\makebox(0,0)[lb]{\raisebox{0pt}[0pt][0pt]{$r$}}}
\put( 75,645){\makebox(0,0)[lb]{\raisebox{0pt}[0pt][0pt]{$a$}}}
\end{picture}$$
	\caption{Open circles and dots represent the original
                         Voisin--Borcea orbifolds.
                         The line of $\otimes$'s corresponds to the extension
                         discussed in the text.}
	\label{figVB}
\end{figure}
%%%%%%%%%%%%%%%%%%%%%%%%%%%%%%%%%%%%%%%%%%%%%%%%%

\end{document}